\documentstyle[aps,prb,floats,twocolumn]{revtex}
\input epsf     

\begin{document}
\draft
\title{Evaluating and improving the cluster variation method
entropy functional for Ising alloys}
\author{Luiz G. Ferreira$^{(a),(b)}$, C. Wolverton$^{(a)}$, and Alex Zunger$^{(a)}$}
\address{$^{(a)}$National Renewable Energy Laboratory, Golden, CO 80401\\}
\address{$^{(b)}$Instituto de F\'{\i}sica, 
Universidade Estadual de Campinas, 13083-970 Campinas, S\~ao Paulo, Brazil}
\date{\today}
\maketitle
%
%
{\let\clearpage\relax
\twocolumn[%
\widetext\leftskip=0.10753\textwidth \rightskip\leftskip
%
\begin{abstract}
The success of the ``Cluster Variation Method'' (CVM) in reproducing
quite accurately the free energies of Monte Carlo (MC) calculations
on Ising models is explained in terms of identifying a cancellation
of errors:  We show that the CVM 
produces correlation functions that are too close to zero, which
leads to an {\em overestimation} of the exact energy, $E$, and
at the same time, to an {\em underestimation} of $-TS$, so the
free energy $F=E-TS$ is more accurate than either of its
parts.  This insight explains a problem with
``hybrid methods'' using
MC correlation functions in the CVM entropy expression:
They give exact energies $E$ and do not give significantly
improved $-TS$ relative to CVM, so they do not benefit from
the above noted cancellation of errors.
Additionally, ``hybrid methods'' suffer from the
difficulty of adequately accounting for both
ordered and disordered phases in a consistent way.
A different technique, the ``Entropic Monte Carlo'' (EMC),
is shown here to provide a means for critically
evaluating the CVM entropy.
Inspired by EMC results, we find a universal and simple
correction to the CVM entropy
which produces individual components of the free energy
with MC accuracy, but is computationally much less 
expensive than either MC thermodynamic integration or EMC.
\end{abstract}
\vspace{11pt}
\pacs{PACS numbers: 64.60.C, 05.50.+q}

]}
\narrowtext

\section{Introduction}

The physics of phase transitions and phase stability of alloys is
often couched in terms of 
statistical mechanics models on the generalized 
(long-range pair and multibody interactions) Ising 
lattice and computed most accurately with Monte Carlo
(MC) methods \cite{Binderbook}. These are time consuming calculations and
usually are thus restricted to coarse grids of chemical
potential and temperatures. Further, MC simulations do not give 
directly the values
of important thermodynamic variables such as entropy and free energy,
since these quantities cannot be written in terms of ensemble averages.
Instead, these are obtained laboriously 
by integration of thermodynamic relations from a known starting point.
To remedy this situation, one often uses the less accurate
Molecular Field methods,
most notably the Cluster Variation Method (CVM) of Kikuchi \cite{Kikuchi}.
Despite its great simplicity, the CVM reproduces 
many of the features of phase diagrams obtained by the
many-orders-of-magnitude more computer expensive MC method: 
For the fcc nearest-neighbor antiferromagnetic Ising
Hamiltonian with coupling constant $J$ and
zero chemical field $h$=0,
the transition temperature $T_c$ \cite{note} 
from MC simulations is 1.74 \cite{tc}
while CVM in the Tetrahedron (Tetrahedron-Octahedron)
approximation gives 1.89 (1.81). 
This and other successes of the CVM \cite{Kikuchi} are 
even more surprising in light of the finding 
that the CVM correlation functions (the thermal
average of products of the Ising spin variables)
differ considerably from the exact MC values:  
For example, in the nearest-neighbor
fcc antiferromagnetic Ising model, the MC pair 
correlation functions \cite{note.mc} at 
$T=1.9$ and $h=0$ are $-0.208,\,\,0.254,\,\,0.036,\,\,0.076$ for the first
to fourth neighbors, respectively,
while the tetrahedron-CVM first neighbor
correlation function is
$-0.188$
and the tetrahedron-octahedron-CVM first and second neighbor
correlation functions are
$-0.198$ and $+0.198$.
Thus, CVM correlation functions
are substantially closer to zero (i.e., more ``random'')
than the exact values. 
The error in tetrahedron-CVM first neighbor
correlation function leads to a $\sim$10$\%$ error in
both energy and entropy relative to MC (see below).
However, despite such systematic discrepancies (of $\sim$10\% or
less) in reproducing 
correlation functions, the CVM seems to describe well thermodynamic
properties (e.g., free energies) which depend on these
very correlation functions.  The subject of this paper is precisely
these types of errors in CVM energy, entropy, and free energy 
relative to MC.  We make four points:

(i) We show that a reason for the success of the CVM
in describing the {\em free energy} 
is an interesting cancellation of errors: 
The closer-to-zero CVM correlation functions imply greater
randomness and hence an {\em overestimation} of
the internal energy compared to MC. 
However, the more random CVM correlations
also lead to a larger entropy, and hence to an
{\em underestimation} of the $-TS$ term.
Thus, the error in internal energy is of opposite sign to the
error in the $-TS$ term, so these two errors partially cancel in
the free energy.  This cancellation of errors is due to
the fact that the CVM free energy expression may be obtained
from a variational argument, \cite{Kikuchi} but not $E$ or
$S$ individually.

(ii) Our analysis gives insight into the successes and failures
of various approaches that attempt to improve CVM by
``borrowing'' certain quantities from MC.  Indeed,
for some applications, one may require accuracies and flexibilities 
beyond those provided by the CVM, and so there is a desire in the field
to find accurate ``hybrid'' methods combining the simplicity of the CVM
with the accuracy and flexibility of MC.  
A natural possibility is to use \cite{Chris}
the correlation functions $\Pi_{\rm MC}$
(or cluster probabilities) obtained
from MC simulations in the 
CVM expression for free energy
$F_{\rm CVM}(\Pi_{\rm MC})$
in the hope of obtaining a more
accurate free energy.
We demonstrate that these methods \cite{Chris} are 
unlikely to succeed, as these
approximations do not benefit from the cancellations of errors noted
above.  

(iii) We use the ``Entropic Monte Carlo'' (EMC) method
of Lee, \cite{Lee}  which provides a method for determining the
entropy as a function of any state variable.  We apply EMC
to the case of the CVM entropy as the state variable, 
and demonstrate that $S_{\rm EMC}(S_{\rm CVM})$
provides a means for critically
evaluating the errors in CVM entropy. 
The calculation of 
$S_{\rm EMC}[S_{\rm CVM}(\Pi_{\rm MC})]$
further shows that this functional accurately
describes $S_{\rm MC}(\Pi_{\rm MC})$
and thus the Monte Carlo free energy.
However, this approach is computationally intensive.
Finally, inspired by the EMC philosophy,

(iv) We develop a functional 
$\tilde{S}[S_{\rm CVM}(\Pi_{\rm MC})]$
that reproduces the exact
$S_{\rm MC}(\Pi_{\rm MC})$
very closely, and is computationally much less
expensive than either MC thermodynamic integration or EMC.
This functional permits one to borrow from MC calculations
the correlation functions (or cluster probabilities),
evaluate the ensuing CVM entropy $S_{\rm CVM}(\Pi_{\rm MC})$,
and thus obtain the nearly exact entropy
$\tilde{S}[S_{\rm CVM}(\Pi_{\rm MC})] \approx
S_{\rm MC}(\Pi_{\rm MC})$ and energy
$E(\Pi_{\rm MC})$.
This approach thus combines the accuracy of MC with
the computational simplicity of the CVM.

\section{Methods}

All of the calculations described in this paper will be
tests of the various methods on the fcc nearest-neighbor 
antiferromagnetic Ising model.

\subsection{CVM Quantities}

We first briefly review our notation. Let $\sigma $ mean a
configuration (``microstate'') of Ising spins ($\pm 1$) on a lattice. 
Consider a cluster (``figure'') $f$ with $k_f$ lattice points. The spin
variable, which 
takes on the value $\hat{S}_i(\sigma) = -1 (+1)$ if there is an A (B)
atom at site $i$ of the figure,
depends on the configuration $\sigma$ of spins in the lattice.
Consider now all the clusters $Rf$
obtained from the cluster $f$ by the symmetry operations $R$ of the space
group of the lattice.  In the CVM, we define the correlation function $\bar{%
\Pi}_f\left( \sigma \right) $ for the cluster $f$ in the configuration 
$\sigma$ as the product of spin variables over the sites of $f$,
averaged over all the figures obtained from $f$ by the space group operations $R$:
\begin{equation}
\bar{\Pi}_f\left( \sigma \right) =
\frac{\sum_R \hat{S}_1 \hat{S}_2 ... \hat{S}_{k_f}}{\sum_R1}.
\label{pi}
\end{equation}
The CVM treats the correlation functions $\{\bar{\Pi}_f\}$ as thermodynamic
variables. For a cluster $f$ of $k_f$ sites there are $2^{k_f}$ arrangements
of spins $\pm 1$ at its sites. Each arrangement $j$ has a cluster
probability $\rho _j^{f}$ which is linearly dependent on the correlation
function values for all subclusters of $f$. 
The correlation functions $\{\bar{\Pi}_f\}$ (or equivalently, cluster probabilities)
are determined by minimizing the free energy,
composed of the CVM internal energy
\begin{equation}
\label{energy}
E_{\rm CVM} = \sum_{f\subseteq F} D_f J_f \langle \bar{\Pi}_f \rangle
\end{equation}
and the CVM entropy 
\begin{equation}
\label{scvm}
S_{\rm CVM}=-k\sum_{f\subseteq F}B^{f}\sum_j\rho _j^{f}\ln (\rho _j^{f})
\end{equation}
both written as a sum over all the subclusters of the maximum $F$.  In the CVM
entropy expression, one also sums 
over the arrangements of spins at the sites
of each subcluster. 
The Barker coefficients 
$B^{f}$ can be obtained from purely group theoretical 
arguments.  \cite{Barker,Morita}
Unless otherwise noted, all CVM calculations described in this
paper are for the fcc tetrahedron approximation.


In order to examine the errors involved in the CVM, we first
compute the accurate energy, entropy, and free energy from 
Monte Carlo simulations of the nearest-neighbor 
anti-ferromagnetic Ising model at $h$=0.  

\subsection{Monte Carlo Quantities}

A Monte Carlo cell of 1728 sites
was used with 10$^6$ Monte Carlo steps per site at each
temperature.  Although finite-size effects were not taken
into account, the calculated heat capacity showed a sharp peak
at $T=1.77$, within $\sim$1-2\% of
the most precise values for the transition temperature
given in the literature $\sim$1.74-1.75. \cite{tc}
The energy is given directly from MC, while the entropy is obtained
by thermodynamic integration down from
infinite temperature:

\begin{equation}
S\left( T\right) =k\ln 2+\frac{E\left( T\right) }{T}-\int_{0}^{\frac{1}{T}%
}E\left( T \right) \; d(1/T)
\end{equation}
(The entropy was also obtained by integrating the heat capacity down
in temperature, however, this method was found to be less efficient
in that it required a finer grid of temperatures near the transition
for equal accuracy).

The correlation functions $\Pi_{\rm MC}$ for a figure $f$
were obtained by taking the thermal average (over the
10$^6$ Monte Carlo steps) of the
product of Ising spin variables over the sites ${1,2,...,k_f}$
of all symmetry-equivalent figures $f$ 
[i.e., the thermal average of Eq. (\ref{pi})].

\section{Results}
\subsection{Analysis of CVM errors vis-a-vis MC simulations}
\label{errors}

%
%
\begin{figure}[tb]
\hbox to \hsize{\epsfxsize=0.80\hsize\hfil\epsfbox{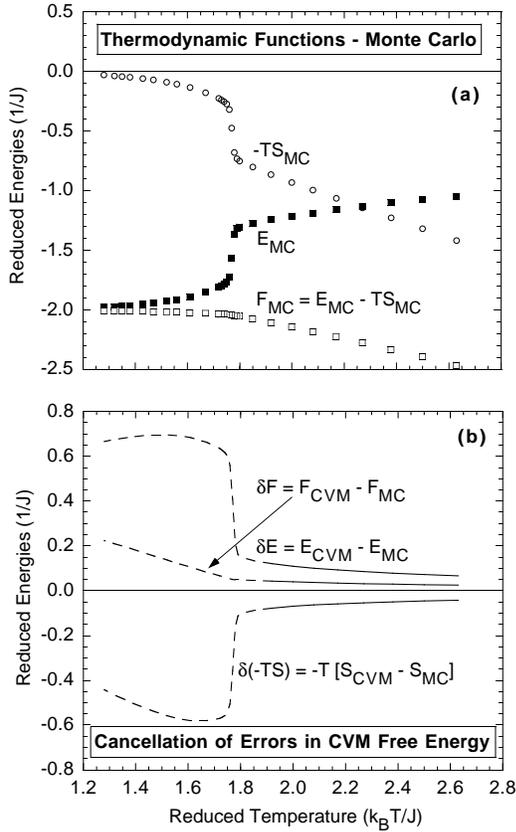}\hfil}
\nobreak\bigskip
\caption{Energy, entropy, and free energy as a function of
temperature for the nearest-neighbor
anti-ferromagnetic Ising model.
All quantities are given in dimensionless units:
$k_BT/J$ for temperature,
energies are given normalized by $J$,
and entropies are given normalized by $k_B$.
(a) Results obtained from
Monte Carlo simulations and thermodynamic integration.
(b) Errors in standard CVM (disordered phase symmetry)
compared to Monte Carlo.
We have used only the CVM entropy expression with the
disordered phase symmetry.  Thus, in (b), differences with
Monte Carlo for temperatures below the CVM transition
($T$=1.89) are overestimated and hence are
shown as dashed lines (see text).
}
\label{mc}
\end{figure}

The energy, entropy, and free energy obtained from Monte Carlo
simulations are shown in Fig. \ref{mc}a.  The first-order transition
at $T \simeq 1.77$ is evident from the discontinuity in energy and
entropy.  We have also computed the energy, entropy, and 
free energy predicted by CVM (in the tetrahedron approximation).
By comparing these CVM results with the ``exact'' Monte Carlo
results in Fig. \ref{mc}a, we may ascertain the errors in
thermodynamic functions of the CVM.  The {\em differences} 
$\delta E$, $\delta (-TS)$, and $\delta F$ between
the respective
CVM and Monte Carlo functions are shown in Fig. \ref{mc}b.
In our CVM calculations,
we have used only the CVM entropy expression with the
disordered phase symmetry.  Thus, in Fig. \ref{mc}b differences with
Monte Carlo for temperatures below the CVM transition
($T$=1.89) are shown as dashed lines.  Obviously,
practitioners of CVM would correctly impose a lower symmetry 
on the entropy expression below the transition to the 
ordered phase and would not use the disordered phase
symmetry in this temperature range.
The reason we use the CVM disordered phase symmetry
down to low temperature is due to our wish to combine
CVM methods with MC, which as we describe in the next
section, is problematic when using ordered expressions
for CVM entropy.

As indicated in the Introduction, the CVM correlation
functions $\{\bar{\Pi}_f\}$ are closer to zero than
the MC values.  By Eq. (\ref{energy}), the CVM internal
energy is less negative
relative to Monte Carlo (thus, $\delta E>0$ in Fig. \ref{mc}b), 
demonstrating that
the energetic effect of short-range order in CVM is 
underestimated, and hence the CVM internal energy is ``more random'' 
than that of Monte Carlo.
The entropy of CVM 
is overestimated ($\delta (-TS)<0$) relative
to Monte Carlo, again indicating a more random solution than
Monte Carlo.  However, the error in the CVM {\em free energy}
$\delta F$
is simply the sum of the errors $\delta E + \delta (-TS)$.  Since
$\delta E$ and $\delta (-TS)$
have opposite sign, they partially cancel, and
give an error in free energy which is considerably
smaller in magnitude than
either the error in energy or in entropy.  Thus, owing to the
variational nature of the CVM, \cite{Kikuchi} the free energy
of CVM is more accurate than one might expect from considering 
either the energy or entropy alone.

\subsection{Using MC correlation functions in CVM calculations:
Absence of cancellation of errors}
\label{fcvm.pimc}

%
%
\begin{figure}[tb]
\hbox to \hsize{\epsfxsize=0.80\hsize\hfil\epsfbox{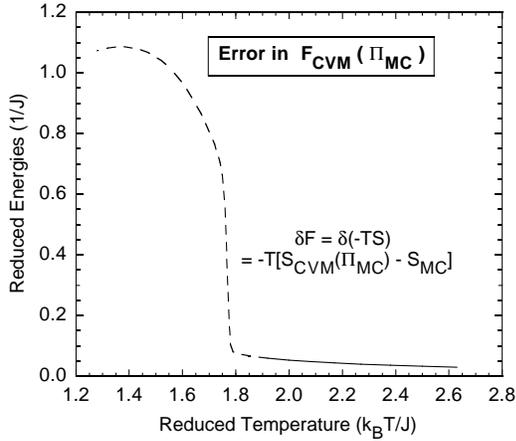}\hfil}
\nobreak\bigskip
\caption{Error in
free energy as a function of
temperature for the nearest-neighbor
anti-ferromagnetic Ising model obtained
using the ``$F_{\rm CVM}(\Pi_{\rm MC})$'' method.
All quantities are given in dimensionless units:
$k_BT/J$ for temperature,
energies are given normalized by $J$,
and entropies are given normalized by $k_B$.
In the ``$F_{\rm CVM}(\Pi_{\rm MC})$''
method, MC correlation functions and
cluster probabilities are used
in the CVM expressions for energy and entropy,
respectively.
We have used only the CVM entropy expression with the
disordered phase symmetry.  Thus, differences with
Monte Carlo for temperatures below the CVM transition
($T$=1.89) are overestimated and hence are
shown as dashed lines (see text).
Note that in this method,
the energy is precisely that of Monte
Carlo, thus the error in $-TS$ is also the error in
free energy.
}
\label{cvm.mc}
\end{figure}

Our foregoing discussion sheds light on a hybrid method which,
naively thinking, might combine the accuracy of Monte Carlo
with the simplicity of CVM.  In this method, one uses the
correlation functions $\{\bar{\Pi}_f\}$ 
(or equivalently, the cluster probabilities $\{\rho _j^{f}\}$)
of MC simulations
in the expressions for CVM entropy [Eq. (\ref{scvm})]
and energy [Eq. (\ref{energy})].  
We refer to this method
as the ``$F_{\rm CVM}(\Pi_{\rm MC})$'' method.  
This method would, of course,
require one to perform a Monte Carlo simulation for each
composition and temperature of interest;  however, one
could, in principle, obtain the entropy at each point
from a {\em single} Monte 
Carlo simulation (i.e., one composition and one temperature)
rather than a {\em series} of Monte Carlo calculations which would be 
required for thermodynamic integration of the entropy.
Since the Monte Carlo correlation functions are
used in this method in Eq. (\ref{energy}), there
is no error in energy ($\delta E=0$).  Thus, the error in
free energy, shown in Fig. \ref{cvm.mc}, 
is equal to the error in entropy:
$\delta F = \delta (-TS)$.
Since there is no error in energy in this method, there is
no cancellation of errors.  Hence, {\em even though
the exact Monte Carlo correlation functions are used
in the ``$F_{\rm CVM}(\Pi_{\rm MC})$'' method, it
produces less accurate free energies than standard
CVM}:  For example, at $T$=1.92, the free energies 
as given by Monte Carlo, ``$F_{\rm CVM}(\Pi_{\rm MC})$'', 
and CVM, are
= -2.109, -2.050, and -2.067, respectively. 
(For comparison, CVM in the Tetrahedron-Octahedron
approximation gives $F$ = -2.094 for this temperature.)

%
%
\begin{figure}[tb]
\hbox to \hsize{\epsfxsize=0.80\hsize\hfil\epsfbox{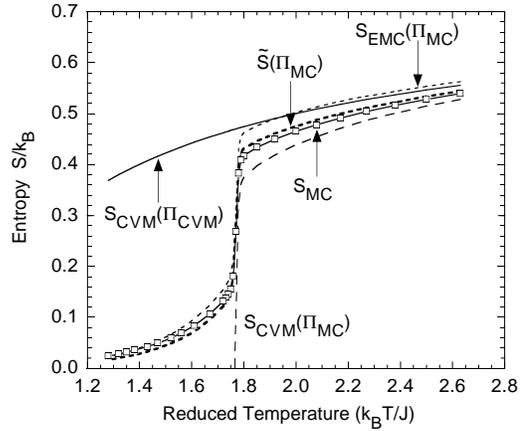}\hfil}
\nobreak\bigskip
\caption{Entropy versus temperature for the
nearest-neighbor anti-ferromagnetic Ising model.
The open squares (connected by a solid line) is
the result of Monte Carlo simulations, the
solid line is standard CVM, the long dashed line is
the CVM entropy expression evaluated with the
Monte Carlo cluster probabilities, the
thin short dashed line is the result of the entropic
Monte Carlo calculations, and
the thick short dashed line is the simple correction
to the CVM, the ``modified CVM''.}
\label{entropy}
\end{figure}

Other disadvantages of the $F_{\rm CVM}(\Pi_{\rm MC})$
approach are illustrated in
Fig. \ref{entropy}, showing a comparison of the
entropies as a function
of temperature as calculated by MC ($S_{\rm MC}$), 
by standard CVM [$S_{\rm CVM}(\Pi_{\rm CVM})$],
and by the $F_{\rm CVM}(\Pi_{\rm MC})$ method
[$S_{\rm CVM}(\Pi_{\rm MC})$].  
(The other two curves of Fig. \ref{entropy} are discussed in
Sections \ref{emc} and \ref{mcvm} below.)
One can again see that
(i) standard CVM (solid line)
overestimates the entropy at high temperatures relative
to Monte Carlo (open squares), 
(ii) the CVM entropy of the disordered phase
is not applicable at low temperatures,
and 
(iii) the $F_{\rm CVM}(\Pi_{\rm MC})$ method underestimates the entropy
at high temperatures, and at low temperatures this entropy 
takes on the unphysical values 
$S_{\rm CVM}(\Pi_{\rm MC})<0$.
These unphysical values are a result of the fact that
the expression for the {\em disordered} $S_{CVM}$ allows
negative values for many atomic configurations. 
For instance, in
the {\em fcc} lattice, the simplest ordered configurations such as 
$L1_0,\,\,L1_1,\,\,L1_2$ have negative $S_{CVM}$ values. 
(These negative values of CVM entropy
are likely to persist no matter what sized maximal cluster
is used, and thus will always lead to difficulties with the
$F_{\rm CVM}(\Pi_{\rm MC})$ method at lower temperatures,
since the CVM entropy will incorrectly tend to negative values 
rather than zero.)
Of course, one might argue that these ordered configurations
only possess negative CVM entropy when evaluated with the
CVM expression for the disordered phase, 
whereas when evaluated with the
CVM expressions for the corresponding ordered phases, 
they will have non-negative
entropies.  However, this illuminates another potential
problem with the $F_{\rm CVM}(\Pi_{\rm MC})$ method:  
In Monte Carlo simulations,
the presence of anti-phase boundaries and finite-sized domains
of long-range order preclude one from unambiguously
defining the distinction between  sublattices and extent of 
long-range order present in the simulation.  
However, an {\em ordered} CVM expression for the entropy 
is written in terms of correlation functions and 
cluster probabilities for all the symmetry-distinct
figures for the symmetry of the long-range ordered
phase.  
Thus, the CVM ordered entropy expressions
presuppose the domains of long-range order are
infinite in size, and hence the distinction of various 
sublattices in the ordered phase is unambiguous.
Thus, using the ``$F_{\rm CVM}(\Pi_{\rm MC})$'' method
with ordered CVM entropy expressions is not
practical because one doesn't know from the MC simulations
precisely how to divide the MC simulations into 
sublattices of long-range order and hence one does
not even know from the MC simulations which ordered
CVM expression to use.
Also one
does not know at what temperature to change the symmetry of
the CVM to the ordered entropy expression.
Thus, an ideal method combining Monte Carlo and CVM would only
use a {\em single} expression (e.g., the disordered CVM expression)
for the entropy at both low and
high temperatures.  
We next describe such a method called ``Entropic
Monte Carlo'' (EMC).

\subsection{The entropic Monte Carlo method:  A critical
evaluation of CVM entropy}
\label{emc}

Lee \cite{Lee} has shown a practical way to determine the entropy of a Monte
Carlo cell as a function of any state variable. We call this method
Entropic Monte Carlo (EMC).  Though in his paper Lee
applies the EMC method
to the case of the energy as state variable in a quantized
system, here we describe instead $S_{\rm EMC}$ in terms
of the state variable 
$S_{CVM}(\sigma)$ which is a continuous, not quantized, variable.
Our strategy will be to calculate $S_{\rm EMC}(S_{CVM})$
by the method of Lee, and then insert $S_{CVM}(\Pi_{\rm MC})$
into this expression, giving $S_{\rm EMC}[S_{CVM}(\Pi_{\rm MC})]$
which we write as $S_{\rm EMC}(\Pi_{\rm MC})$.  We will show that
this function reproduces very well $S_{\rm MC}$.
First, we describe how $S_{\rm EMC}(S_{CVM})$ is calculated:

The EMC method is a self-consistent process in which each iteration 
is made from a
series of Monte Carlo sweeps where the driving ``energy'' $E(\sigma)$ 
of the Monte Carlo equations is not the true energy contained in the sample 
but an approximation to the {\em entropy}:
\begin{equation}
E(\sigma) =E[S_{CVM}(\sigma)]
\label{driving}
\end{equation}
which depends on the configuration $\sigma $ through the 
function $E[S_{CVM}] $, whose argument is the CVM entropy (per site) 
calculated with Eq.(\ref{scvm})
for the cluster probabilities $\rho _j^{f}\left( \sigma \right) $ of the
configuration $\sigma $. The function $E\left[ S_{CVM}\right] $ is assumed
to be monotonic. 
The EMC dynamics are given by the detailed balance condition 
\begin{equation}
\exp \left[ -E\left( \sigma _i\right) \right] W(i\rightarrow j)=\exp \left[
-E\left( \sigma _j\right) \right] W(j\rightarrow i)
\end{equation}
where $W$ are transition rates and
$E(\sigma)$ is given by Eq. (\ref{driving}). 
After many MC sweeps of the full lattice, one 
obtains a histogram 
\begin{equation}
H\left( \bar{S}_{CVM}\right) =Xd\left( \bar{S}_{CVM}\right) \exp \left[
-E\left( \bar{S}_{CVM}\right) \right]
\end{equation}
where $d\left( \bar{S}_{CVM}\right) $ is the number of configurations with a
given value $\bar{S}_{CVM}$ of the CVM entropy (degeneracy in $S_{CVM}$) and 
$X$ is a constant of proportionality. We distinguish $\bar{S}_{CVM}$, which
is a numerical argument attaining certain value, from $S_{CVM}$ which is a
function of both the microstate [through Eq.(\ref{scvm})] and the cluster
probabilities $\rho _j^{f}\left( \sigma \right)$. 
The density of
states function of the CVM entropy $S_{CVM}$ is given by 
\begin{eqnarray}
\label{D}
D(\bar{S}_{CVM}) &=&\sum_\sigma d\left( \bar{S}_{CVM}\right) \delta \left( 
\bar{S}_{CVM}-S_{CVM}\left( \sigma \right) \right)  \nonumber  \\
&=&(1/X)\sum_\sigma \exp \left( E\left( \bar{S}_{CVM}\right) \right) H\left( 
\bar{S}_{CVM}\right) \times \nonumber \\
&& \delta \left( \bar{S}_{CVM}-S_{CVM}\left( \sigma
\right) \right)  
\end{eqnarray}
and the entropy $S_{EMC}=S(\bar{S}_{CVM})$ per site is defined as 
\begin{equation}
\exp [NS(\bar{S}_{CVM})]=\int_{-\infty }^{\bar{S}_{CVM}}D(\xi )d\xi
\label{defSS}
\end{equation}
\noindent where $N$ is the number of sites in the MC cell.

From Eqs.(\ref{D}) and Eq.(\ref{defSS}), we may obtain the difference between 
the entropy at two different values $\bar{S}_{CVM}$.
\begin{eqnarray}
S(&&\bar{S}_{CVM}^{(2)})-S(\bar{S}_{CVM}^{(1)})= \nonumber \\
& &\frac{1}{N}\ln [
\sum_{\sigma ,\,\,S_{CVM}(\sigma )<\bar{S}_{CVM}^{(2)}}
\exp(NE[S_{CVM}(\sigma)]) \times \nonumber \\
& &H[S_{CVM}(\sigma)]]   \nonumber \\
& &-\frac{1}{N}\ln [\sum_{\sigma ,\,\,S_{CVM}(\sigma) <
\bar{S}_{CVM}^{(1)}}\exp(NE[S_{CVM}(\sigma)]) \times \nonumber \\
& &H[S_{CVM}(\sigma)]]
\label{basic}
\end{eqnarray}
which is the basic equation used to determine the entropy from the EMC runs.
On the right-hand side of Eq. (\ref{basic}), the sums are over
the microstates $\sigma$ obtained in the EMC sweeps whose CVM entropy
$S_{CVM}(\sigma)$ are smaller than $\bar{S}_{CVM}^{(2)}$ or
$\bar{S}_{CVM}^{(1)}$.
As pointed out by Lee \cite{Lee}, the entropy determination becomes
especially simple when the interaction $E(S_{CVM})$ is such that the
histogram $H\left[ \bar{S}_{CVM}\right] $ is uniformly distributed and
has little dependence on $\bar{S}_{CVM}$. In this case, Eq. (\ref{basic}) 
becomes 
\begin{eqnarray}
S(&&\bar{S}_{CVM}^{(2)})-S(\bar{S}_{CVM}^{(1)}) = \nonumber \\
&& \frac 1N\ln [\sum_{\sigma ,\,\,S_{CVM}(\sigma) <\bar{S}_{CVM}^{(2)}}
\exp(NE[S_{CVM}(\sigma)])] \nonumber \\
&&-\frac 1N\ln [\sum_{\sigma ,\,\,S_{CVM}(\sigma) <\bar{S}_{CVM}^{(1)}}
\exp(NE[S_{CVM}(\sigma)])].
\label{basic.simple}
\end{eqnarray}
Because of the factor of $N$, the
exponential in Eq. (\ref{basic.simple})
is a rapidly increasing function of $S_{CVM}$, and hence
only the extremes contribute significantly to the sums, or 
\begin{equation}
S(\bar{S}_{CVM}^{(2)})-S(\bar{S}_{CVM}^{(1)})\simeq \frac 1NE(\bar{S}%
_{CVM}^{(2)})-\frac 1NE(\bar{S}_{CVM}^{(1)})  \label{E}
\end{equation}
This equation 
then suggests the self-consistent procedure for determining the
entropy: 
From a crude estimate of $S\left( S_{CVM}\right) $, 
we use Eq. (\ref{E}) to obtain the interaction 
$E\left( S_{CVM}\right)$ with which we make EMC runs, with which we
recalculate $S\left( S_{CVM}\right) $ from the basic Eq. (\ref{basic}). This
process is taken to self-consistency. 
When self-consistency is reached, 
(i) the histogram 
$H[S_{CVM}]$ (the number of microstates in each small
range of $S_{CVM}$ obtained in a series of MC sweeps) is nearly
constant,  independent of the value of $S_{CVM}$ and
(ii) the driving ``energy'' which is the
approximation to the entropy [Eq. (\ref{driving})], 
becomes equal to the entropy calculated from
the density of states [Eq. (\ref{defSS})], 
or in other words, the EMC entropy becomes exact. 
An important aspect of the EMC is that the 
calculated entropy functional form of $S\left( S_{CVM}\right) $ {\em does not
depend on any particular Ising Hamiltonian}, (so long as the important
correlations are contained within the CVM maximal cluster),
because the role of the ``energy'' driving the EMC calculations
is played in Eq. (\ref{basic}) by the entropy itself.

%
%
\begin{figure}[tb]
\hbox to \hsize{\epsfxsize=0.80\hsize\hfil\epsfbox{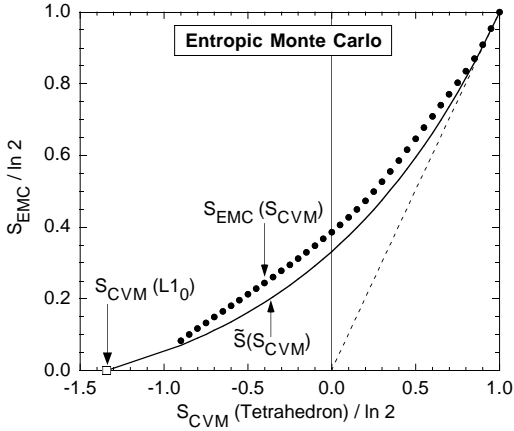}\hfil}
\nobreak\bigskip
\caption{Entropic Monte Carlo results for
$S_{\rm EMC} = S(S_{\rm CVM})$.
Filled circles are the EMC calculations, and
the solid line is the EMC-inspired entropy
functional $\tilde{S}(S_{\rm CVM})$.  The
dashed line is line of unit slope simply to guide
the eye.  EMC was performed for
a cell of 12$^3$=1728 sites, using the CVM tetrahedron
expression for the disordered entropy.  Note that
many configurations correspond to negative CVM
entropy, with the most
negative (for all configurations with $\leq$16 atoms
per cell) being the $L1_0$ configuration.}
\label{old.friend}
\end{figure}

Figure \ref{old.friend} shows a typical result $S_{\rm EMC}(S_{\rm CVM})$
of the EMC calculations,
using a MC cell with $N=12^3=1728$ sites. 
We also performed EMC calculations with
different MC cells, with the results being slightly different for the larger
negative values of $S_{CVM}$. The negative values of $S_{CVM}$ correspond to
configurations of atoms with higher symmetry, usually associated with
smaller repeat units, thus explaining why the curve depends to some
extent on the size and shape of the MC cell for the negative values of 
$S_{CVM}$: For instance, for an EMC cell with an odd number of sites (e.g., 
$11^3$), one could never obtain the stoichiometric
configuration $L1_{0}$ with its large negative CVM entropy. In fact, simple
high-symmetry configurations such as $L1_{0},\,L1_{1},\,L1_{2}$ all have 
negative values of $S_{CVM}$. Examining the CVM entropy for all 
configurations with up to 16 atoms per cell, \cite{file} 
we found the configurations with the 
most negative CVM entropy 
had very small unit cells.  The largest negative
CVM entropy occurs for 
$L1_{0}$ for which $S_{CVM}\left( L1_{0}\right) =-1.34\ln 2$. 

The results $S_{\rm EMC}[S_{\rm CVM}(\Pi_{\rm MC})]$
of EMC are shown in Fig. \ref{entropy}, where they are
contrasted with the results of Monte Carlo, 
CVM, and ``$F_{\rm CVM}(\Pi_{\rm MC})$''. \cite{note.emc}
By comparing $S_{\rm EMC}(\Pi_{\rm MC})$ and
$S_{\rm CVM}(\Pi_{\rm CVM})$ with $S_{\rm MC}$,
we see that
the EMC and CVM entropies are equally accurate 
at high temperatures.
Remarkably however, 
the EMC method also
produces extremely accurate entropies at 
{\em low temperatures}, in qualitative contrast with
the ``$F_{\rm CVM}(\Pi_{\rm MC})$''
method.  Thus, even though
one only uses a single disordered expression for
the CVM entropy in the EMC calculations, the EMC
reproduces both high temperature (disordered) 
and low temperature (ordered)
entropy values, with no need
to change the CVM entropy expression at any point.
Although the internal energy in EMC is exact (so this
method does not benefit from the cancellation of errors
noted in Sec. \ref{errors} for the CVM), we see that EMC
does not need to be correct due to {\em cancellation}
of errors.  Instead, it is accurate because its
{\em individual terms} ($E$ and $-TS$) are accurate.

The EMC, like standard Monte Carlo, can be a computationally laborious
procedure.  However, our EMC calculations of $S\left( S_{CVM}\right)$ 
suggest a very simple functional $\tilde{S}(S_{\rm CVM})$
which is appealing
because the correction {\em does not require one to perform an EMC
calculation}.  We next describe this simple correction.

\subsection{An EMC-inspired new entropy functional}
\label{mcvm}

While $S_{\rm CVM}(\Pi_{\rm MC})$ can be inaccurate,
$S_{\rm EMC}[S_{\rm CVM}(\Pi_{\rm MC})]$ is
accurate but computationally expensive.  Thus we will now
develop a new functional $\tilde{S}[S_{\rm CVM}(\Pi_{\rm MC})]$
which is both accurate and inexpensive.

The EMC results of Fig. \ref{old.friend}
permit one to guess the behavior of the ``exact'' entropy $S\left(
S_{CVM}\right) $ (in the limit of $N\rightarrow \infty $)
as a function of the CVM entropy
obtained from a ``good'' maximal cluster (e.g., the tetrahedron or the
tetrahedron-octahedron). 
This ``true'' entropy function $S(S_{CVM})$ should have the following properties:

(i)  The most positive $S_{\rm CVM}(x)$ entropy $S_{\rm CVM}^{\rm MAX}(x)$
should correspond to the exact entropy for this case, i.e., the ideal
mixing entropy:  
$S(S_{\rm CVM}^{\rm MAX})=S^0(x)= -k_B[x\ln x + (1-x)\ln (1-x)]$.

(ii) The slope of $S(S_{CVM})$ at the maximum value of $S_{\rm CVM}^{\rm MAX}=S^0$
should be unity because for nearly random configurations 
the CVM approaches the exact result:
$\frac{dS}{dS_{CVM}}|_{S^0}=1$. 

(iii) The most negative value of the CVM
entropy should correspond to zero ``true'' entropy.
Thus, $S(S_{\rm CVM}^{\rm MIN})=0$. 
The configuration with most negative CVM entropy can be
found by examining all configurations up to some maximum unit-cell size,
as described in Ref. \onlinecite{file}.
For instance, for the tetrahedron CVM,
$L1_{0}$ has the most negative CVM entropy
[$S_{\rm CVM}^{\rm MIN}=S_{CVM}(L1_0) =-1.34\ln 2$].
This point is indicated in Fig. \ref{old.friend} by a square.

(iv) The function $S\left( S_{CVM}\right) $ should increase 
monotonically with $S_{CVM}$ as one can see from Eq. (\ref{basic}). 
Also, $S\left( S_{CVM}\right) $ has a
positive curvature [due to the exponent of $N$ in the 
right-hand side of Eq. (\ref{basic})].

We select a simple functional form for $S\left( S_{CVM}\right) $
which satisfies (i)-(iv) above (but otherwise possesses
no special physical meansing.)  However, use of this
simple form will provide a means of evaluating energies
and entropies which (a) is computationally much more efficient
than either MC thermodynamic integration or EMC, (b) possesses
MC accuracy, and (c) may be extended to use any maximal cluster
of the CVM.
The functional form we choose for the approximation 
$\tilde{S}(S_{\rm CVM})$
to the true $S\left( S_{CVM}\right) $ which satisfies the
four properties (i)-(iv) above is:
\begin{equation}
\tilde{S}(S_{\rm CVM})=
(S^0-\frac{S^0}{\alpha})+
\frac{S^0}{\alpha}
\exp[\alpha(\frac{S_{\rm CVM}}{S^0}-1)] 
\label{MCVM}
\end{equation}
where $\alpha$ is the solution of
\begin{equation}
0=(S^0-\frac{S^0}{\alpha})+
\frac{S^0}{\alpha}
\exp[\alpha(\frac{S_{\rm CVM}^{\rm MIN}}{S^0}-1)].
\label{MCVM.alpha}
\end{equation}
In the case of the tetrahedron CVM, $\alpha =0.86917$. 
The function in Eq. (\ref{MCVM.alpha}) depends only
on a single parameter $S_{\rm CVM}^{\rm MIN}$ which can be easily
estimated from an enumeration of small-unit-cell configurations \cite{file} 
{\em using any maximal cluster of the
CVM method}.  (In contrast, an EMC calculation as we described in
Section \ref{emc} is only practical for the tetrahedron CVM approximation.)
While the negative $S_{CVM}$ configurations, which correspond to
highly symmetric arrangements of atoms, have no meaning in the standard CVM
procedure (since in standard CVM one would use a different expression
for CVM entropy to describe ordered phases), 
the main merit of a correction such as 
Eqs. (\ref{MCVM})-(\ref{MCVM.alpha}) is to
restore these highly ordered configurations into a single CVM 
expression by attributing a
non-negative entropy to them. Naturally this correction, when used together with
the Monte Carlo correlation functions, will be especially important near the
transition when the ordered configurations begin to be important.

To test these ideas, we calculated 
$\tilde{S}[S_{\rm CVM}(\Pi_{\rm MC})] \equiv \tilde{S}(\Pi_{\rm MC})$,
where $\tilde{S}$ is given by Eq. (\ref{MCVM}), and the CVM
is executed within the tetrahedron approximation.
One sees (Fig. \ref{entropy})
that this approach presents a remarkable improvement over the 
$F_{\rm CVM}(\Pi_{\rm MC})$ method in all temperature ranges, 
especially below the transition temperature where 
$S_{\rm CVM}(\Pi_{\rm MC})$ is negative.
Also, the simple functional 
represented by $\tilde{S}(\Pi_{\rm MC})$ effectively
retains all of the improvements over $F_{\rm CVM}(\Pi_{\rm MC})$
that were obtained by the full EMC calculation.  In fact, for high
temperatures, the $\tilde{S}(S_{\rm CVM})$ approach is even closer to the
exact Monte Carlo results than the EMC calculations on which it
was based!  This fact can be understood by examining the EMC calculations
in Fig. \ref{old.friend}:  Figure \ref{old.friend} 
shows that the EMC calculations
using a cell of 1728 sites do not reproduce property (ii) above,
$\frac{dS}{dS_{CVM}}|_{S^0}=1$.  In fact, the slope of the EMC
curve in Fig. \ref{old.friend} is only about 0.82 at the maximum 
value of the entropy.  EMC simulations with even smaller cells were
typically found to possess even smaller slopes.  Presumably, for larger
EMC simulations one would approach the correct slope of unity.
The slope of EMC being smaller than unity means that as one comes down 
from infinite temperature, or $S_{CVM} = \ln 2$, the EMC entropy
maintains a larger value than it should.  This explains why
$S_{EMC} > S_{MC}$ for temperatures above the critical temperature.
Because the $\tilde{S}(S_{\rm CVM})$ approach
was constructed to obey the requirement
$\frac{dS}{dS_{CVM}}|_{S^0}=1$, it corrects the error in the
slope of EMC caused by the finite-sized simulation cell, and hence
improves the entropy above the transition.
The reason that the derivative of EMC is less than one for small
cell sizes is due to the negative $S_{CVM}$ configurations, which
are typically high-symmetry, small-unit-cell configurations.  Thus,
these negative $S_{CVM}$ states are represented more in small EMC
cells relative to configurations with large-unit-cells and low symmetry.
If the density of states in Eq. (\ref{defSS}) becomes artificially
large for the negative region of $S_{CVM}$ (due to small EMC cells), 
it will have to be compensated by an artificially small density
of states in the region of positive $S_{CVM}$ [since the integral
in Eq. (\ref{defSS}) is constrained by the fact that it must be
$2^N$ at $S_{CVM} = \ln 2$.]  Thus, the integral will grow more
slowly than it should for CVM entropies approaching $\ln 2$, and
hence the slope will be less than one.

\section{Summary}

The main accomplishment of this paper is to suggest a simple
functional $\tilde{S}(S_{\rm CVM})$ [Eqs. (\ref{MCVM})-(\ref{MCVM.alpha})]
that improves the CVM entropy.
The development was
based on insights gained from our analysis of the CVM free energy
(which showed cancellation of energetic {\em vs.} entropic errors),
and from the EMC philosophy \cite{Lee} permitting one to express
the true entropy as a functional of an approximate, but deterministic
entropy.  The new functional $\tilde{S}(S_{\rm CVM})$ can be used
in future applications either with CVM alone 
[simply by replacing the CVM
entropy with Eq. (\ref{MCVM}) in any existing CVM program], or
with a combination of CVM and $\Pi_{\rm MC}$ (using
$\tilde{S}[S_{\rm CVM}(\Pi_{\rm MC})]$
as described in this paper).

\begin{center}
{\bf Acknowledgements}
\end{center}

LGF acknowledges support from NREL during his visit, where
much of this work was carried out.
Work at NREL was supported by the Office of Energy Research
(OER) [Division of Materials Science of the Office of Basic Energy
Sciences (BES)], U. S.  Department of Energy, under contract No.
DE-AC36-83CH10093.


\begin{references}
\bibitem{note}		Throughout this paper, all quantities are
			given in dimensionless units:
			$k_BT/J$ for temperature, 
			energies are given normalized by $J$,
			and entropies are given normalized by $k_B$.
\bibitem{note.mc}	These correlations are for a 16$^3$ Monte Carlo
			cell with 500 Monte Carlo steps (per site)
			discarded, and averages taken over the
			subsequent 30,000 steps.
\bibitem{note.emc}	The results presented in Fig. \protect\ref{old.friend} 
			correspond to a 
			grand canonical EMC
			for which the chemical potential, 
			rather than composition $x$ 
			remained fixed. We also made canonical EMC runs by
			flipping pairs of opposite spins, thus maintaining 
			constant $x$ at the
			values 0.50 and 0.25. 
			We obtain very similar results with either 
			the canonical EMC or grand canonical EMC, provided
			that one normalizes both the EMC and CVM 
			entropy with the ideal entropy,
			$-x\ln x-(1-x) \ln (1-x)$,
			instead of $\ln 2$ as in 
			Fig. \protect\ref{old.friend}.
\end{references}
\end{document}